\documentclass[prb,showpacs,twocolumn,aps,superscriptaddress,a4paper]{revtex4-1}
\usepackage{dcolumn,amssymb,amsmath,amsfonts,graphicx,latexsym,color}
\usepackage{epstopdf}
\begin{document}

\title{Novel mobility edges in the off-diagonal disordered tight-binding models}
\author{Tong Liu}
\affiliation{Department of Physics, Southeast University, Nanjing 211189, China}

\author{Hao Guo}
\thanks{Corresponding author: guohao.ph@seu.edu.cn}
\affiliation{Department of Physics, Southeast University, Nanjing 211189, China}

\date{\today}

\begin{abstract}
We study the one-dimensional tight-binding models which include a slowly varying, incommensurate off-diagonal modulation on the hopping amplitude.
Interestingly, we find that the mobility edges can appear only when this off-diagonal (hopping) disorder is included in the model, which is different from the known results induced by the diagonal disorder. We further study the situation where the off-diagonal and diagonal disorder terms (the incommensurate potential) are both included and find that the locations of mobility edges change significantly and the varying trend of the
mobility edge becomes nonsmooth. We first identify the exact expressions of mobility edges of both models by using asymptotic heuristic argument, and then verify the conclusions by utilizing several numerical diagnostic techniques, including the inverse participation ratio, the density of states and the Lyapunov exponent.
This result will perspectives for future investigations on the mobility edge in low dimensional correlated disordered system.
\end{abstract}

\pacs{71.23.An, 71.23.Ft, 05.70.Jk}
\maketitle

\section{Introduction}
\label{n1}
Anderson localization~\cite{1an}, a phenomenon in which the destructive interference prevents wave-propagation in disordered medium, is an active research subject in condensed matter physics. The effect of the spacial dimension on Anderson localization is significant. The scaling theory~\cite{2scal} predicts that there is no localization transition in one and two dimensional systems where all wave functions, at least in the absence of
interactions, are exponentially localized no matter how weak the strength of uncorrelated disorder is. Compared with low dimensional systems, the three-dimensional (3D) Anderson model is very unique because the transition can occur at a finite disorder strength and there exists an energy border separating the localized and extended energy levels. Although this universal behaviour is now well understood, the direct experimental observation of the critical energy, dubbed the mobility edge~\cite{3mott}, remains a challenge due to the hardness to realize the 3D uncorrelated disorder. Thus low dimensional correlated disorder systems attract a lot of research interests. Due to the development of ultracold atomic experiments an optical speckle disorder potential can be realized by projecting a laser beam through a ground glass, and recently the mobility edge trajectory has been determined in the speckle disordered system with sufficiently high energy resolution~\cite{4Kon,5Jen,6Sem}.

Low dimensional quasiperiodic systems~\cite{7Mo2,8Sc1,9roati,10Mo,11Mo,dei,17cai,18wang} which are viewed as the highly correlated disorder can also host mobility edges. As an important paradigm, the Aubry-Andr\'{e} (AA) model~\cite{12aubry}, 1D
tight-binding model with an incommensurate potential has a self-dual symmetry and can undergo a localization transition from the extended to localized phase depending on the relative strength of the incommensurate potential and the hopping amplitude. A very recent work~\cite{13lu,Li} has experimentally investigated the localization properties of a bichromatic incommensurate lattice (a sketchy AA model) and demonstrated that there exists the mobility edge separating the localized and extended states in the intermediate phase.
In the aspect of theory, a unique class of systems with very slowly varying incommensurate potential in real space are revealed by Sarma et.al.~\cite{14sarma}. This deterministic (diagonal) potential is neither random nor simply incommensurate, and there are two mobility edges located at the spectrum where the eigenstates at the band center are all extended whereas the band-edge states are all localized.
Later, different variations~\cite{16PRL,15liu,16liu,Gop,Pu,Tri,Pou} of the 1D AA model containing mobility edges have been studied. By including a long-range
hopping term or constructing special forms of the on-site incommensurate potential, Ref.~\cite{19bid}and Ref.~\cite{20gan} demonstrated the existence of the mobility edge in a certain class of quasiperiodic systems which can be precisely addressed by the self-dual symmetry.

In general, mobility edges in these models are all induced by the diagonal disorder. In this paper, we introduce a class of off-diagonal disordered tight-binding models, which are different from the previous diagonal disordered models since the generating mechanism of mobility edges in the spectrum significantly changes. Our generalized AA model, including a slowly varying, incommensurate off-diagonal modulation on the hopping term, is expressed as,
\begin{equation}\label{tb1}
    \hat H=-\sum_{i=1}^{L-1}(t+\lambda_{i})(\hat{c}_{i}^\dag \hat{c}_{i+1}+h.c.)+\sum_{i=1}^{L}\mu_{i} \hat{n}_{i},
\end{equation}
where $\hat{c}_{i}$ ($\hat{c}_{i}^\dagger$) is the fermionic annihilation (creation) operator, $\hat{n}_i=\hat{c}^\dagger_{i}\hat{c}_{i}$ is the particle number operator,
$L$ is the total number of sites, $\lambda_{i}=\lambda\cos(2\pi\alpha{i^{v}}+\phi_\lambda)$ with $0<v<1$ and $\lambda>0$ being the strength of the incommensurate modulation on the off-diagonal hopping amplitude, and $\mu_{i}=\mu\cos(2\pi\beta{i^{v}}+\phi_\mu)$ with $0<v<1$ and $\mu\ge 0$ being the strength of the on-site potential. Without loss of generality, we choose the parameters $\alpha=(\sqrt{5}-1)/2$, the phase in the modulation $\phi_\lambda =\phi_\mu = 0$ and $v = 0.5$. For convenience, $t = 1$ is set as the energy unit. In particular, when $\beta=0$ or $\beta=(\sqrt{5}-1)/2$, the on-site potential $\mu_{i}$ becomes a constant or slowly varying incommensurate term respectively. In the next section, we will investigate both of such situations, and show that the locations of mobility edges in the presence of constant and slowly varying incommensurate potentials are completely different.

The rest of the paper is organized as follows. In Sec.~\ref{n2}, we theoretically
give the semi-analytical arguments for the Hamiltonian~(\ref{tb1}).
In Sec.~\ref{n3}, we present our numerical results and compare them with the theoretical analysis. The conclusion is summarized in Sec.~\ref{n4}.

\section{semi-analytical methods}
\label{n2}
\begin{figure}
  \centering
  \includegraphics[width=0.5\textwidth]{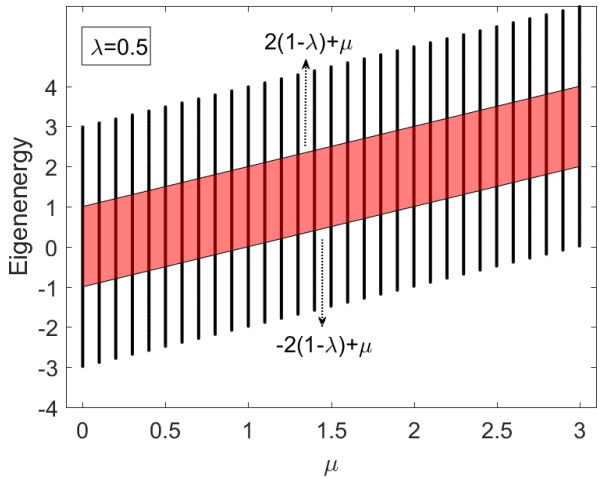}\\
  \caption{(Color online) Eigenenergy of Eq.~(\ref{tb1}) as a function of $\mu$ with $\lambda=0.5$ when $\beta=0$. The total number of sites is set to be $L=10000$ hereafter in this paper. The black curves denote eigenenergies of the system, and the curves covered by the red quadrilateral  correspond to the extended states, whereas uncovered curves correspond to the localized states. The upper and lower solid lines of the red quadrilateral represent two mobility edges denoted by $E_{\pm c1}=\pm 2(1-\lambda)+\mu$. This model does not have an all-wave-function-localized phase transition point.}
  \label{0011}
\end{figure}
\begin{figure}
  \centering
  \includegraphics[width=0.5\textwidth]{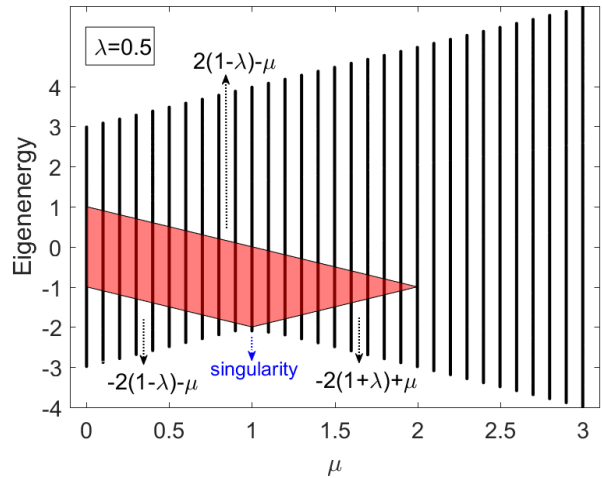}\\
  \caption{(Color online)  Eigenenergy of Eq.~(\ref{tb1}) as a function of $\mu$ with $\lambda=0.5$ when $\beta=(\sqrt{5}-1)/2$. The black curves denote eigenenergies of the system, and the curves covered by the red quadrilateral  correspond to the extended states, whereas uncovered curves correspond to the localized states. The upper and lower solid lines of the red quadrilateral represent two mobility edges denoted by $E_{\pm c2}=\pm 2(1-\lambda)-\mu$ ($0\leq\mu\leq2\lambda$) or $E_{+ c2}= 2(1-\lambda)-\mu$ and $E_{- c2}= -2(1+\lambda)+\mu$ ($2\lambda<\mu<2$). The blue arrow signals the nonsmooth singularity point $\mu=2\lambda$. In this model the all-wave-function-localized phase transition point is located at $\mu=2$.}
  \label{0012}
\end{figure}

The slowly varying incommensurate modulation in 1D system brings up new perspectives to the localization phenomenon due to its highly correlated disorder feature. To demonstrate the existence of mobility edges in Sarma's model~\cite{14sarma}, authors develop an asymptotic semiclassical WKB technique to calculate the density of states and the Lyapunov exponent, and find that these two physical quantities are not necessarily smooth when crossing the mobility edge, which is substantially different from the 3D Anderson model. Here we present some preliminary heuristic arguments to obtain the explicit expressions of mobility edges semi-analytically, then verify our predictions by utilizing several typical numerical techniques.

By noticing that the difference of the off-diagonal slowly varying incommensurate modulation $\lambda_{i}=\lambda\cos(2\pi\alpha{i^{v}})$
vanishes in the thermodynamic limit~\cite{14sarma}, we thus write
\begin{equation}\label{tb2}
\frac{d\lambda_{i}}{di}=-2\lambda\pi\alpha i^{v-1}\sin(2\pi\alpha{i^{v}}).
\end{equation}

When $ i \rightarrow \infty$, Eq.~(\ref{tb2}) can be written as
\begin{equation}\label{tb3}
\lim_{i\rightarrow\infty} \left\vert\frac{d\lambda_{i}}{di}\right\vert=-\lim_{i\rightarrow\infty}  2\lambda\pi\alpha \frac{|\sin(2\pi\alpha{i^{v}})|}{i^{1-v}}=0,
\end{equation}
since $0<v<1$. Equivalently, the modulation difference $ \lambda_{i+1} - \lambda_{i}$ approaches 0 when the lattice number $i$ is large enough, i.e., $\lambda_{i}$ becomes a constant. This asymptotic property of ``being constant" of $\lambda_{i}$ is crucial for the localization
property of our model.

When $\beta=0$, $\mu_{i}$ become a constant $\mu$, Hamiltonian~(\ref{tb1}) can be rewritten as the tight-binding Schr\"{o}dinger equation,
\begin{equation}\label{tb4}
(E-\mu) \psi_{m} + (1+C) \psi_{m-1} + (1+C)\psi_{m+1}= 0,
\end{equation}
where $m$ is an arbitrary positive integer and $C=\lambda\cos(2\pi\alpha{(i-1)^{v}})=\lambda\cos(2\pi\alpha{(i+1)^{v}})$ since all the $\lambda_{i}$ are constants in the large $i$ limit.

From Eq.~(\ref{tb4}) we obtain
\begin{equation}
\psi_{m+1} + \frac{E-\mu}{1+C}\psi_{m}+ \psi_{m-1}= 0.
\label{tb5}
\end{equation}
Following the asymptotic heuristic argument~\cite{14sarma}, we write $\psi_{m}\sim Z^{m}$, where $Z$ is a complex quantity. Then Eq.~(\ref{tb5}) becomes
\begin{equation}
Z^2 + \frac{E-\mu}{1+C}Z+ 1= 0,
\label{tb6}
\end{equation}
to which the complex solutions are
\begin{equation}
Z_{1,2}=\frac{-B\pm\sqrt{B^2-4}}{2},
\label{tb7}
\end{equation}
with $B=\frac{E-\mu}{1+C}$. From Eq.~(\ref{tb7}) we conclude that the amplitude $Z$ is complex/extended if $|B|<2$ whereas real/localized if $|B|>2$. Note $(1+C)_\textrm{min}=1-\lambda$, due to the fact that $\mu$ is a constant number, we can incorporate $\mu$ into $E$ and get
\begin{equation}\label{tb8}
B_\textrm{max} = \frac{|E-\mu|}{1-\lambda}.
\end{equation}
Therefore the conditions for extended and localized solutions are respectively given by
\begin{equation}\label{tb9}
\begin{split}
    & B_\textrm{max}<2 \Rightarrow -2(1-\lambda)+\mu<E<2(1-\lambda)+\mu\\
    & (\textrm{extended}),\\
    & B_\textrm{max}>2 \Rightarrow E<-2(1-\lambda)+\mu \cup 2(1-\lambda)+\mu<E\\
    & (\textrm{localized}).
\end{split}
\end{equation}

Note that for the existence of the mobility edges there is an implicit condition such that $1-\lambda>0$. If this condition is satisfied there will be two mobility edges denoted by $ E_{\pm c1}=\pm 2(1-\lambda)+\mu$. That is, the model defined by the Hamiltonian~(\ref{tb1}) with
$\beta=0$ has extended states at the band center ($E_{-c1}<E<E_{+c1}$) and localized states at the band edges ($E>E_{+c1}$, $E<E_{-c1}$). Interestingly, even if $\mu$ becomes very large, the mobility edges $ E_{\pm c1}$ still exist, as shown in Fig.~\ref{0011}. This is the new feature of the model with off-diagonal slowly varying incommensurate modulation,
whereas the previous models with diagonal slowly varying incommensurate modulation always drives the system to a all-wave-function-localized transition point and the mobility edge vanishes accordingly.

When $\beta=(\sqrt{5}-1)/2$, the system becomes more complicated. In this case $\mu_{i}$ denotes the slowly varying incommensurate potential, thus there exist competing off-diagonal and diagonal incommensurate terms in Hamiltonian~(\ref{tb1}), which can be written as
\begin{equation}\label{tb41}
(E-C') \psi_{m} + (1+C) \psi_{m-1} + (1+C)\psi_{m+1}= 0,
\end{equation}
where $C'=\mu\cos(2\pi\beta{i^{v}})$, other parameters are the same as these in Eq.~(\ref{tb4}).
Following the same procedure, we can obtain $B'=\frac{E-C'}{1+C} $. Note $(1+C)_\textrm{min}=1-\lambda$ and $(E-C')_\textrm{max}=|E+\mu|$, then
\begin{equation}\label{tb81}
B'_\textrm{max} = \frac{|E+\mu|}{1-\lambda},
\end{equation}
Thus, the conditions for extended and localized solutions are respectively given by
\begin{equation}\label{tb91}
\begin{split}
    & B'_\textrm{max}<2 \Rightarrow -2(1-\lambda)-\mu<E<2(1-\lambda)-\mu\\
    & (\textrm{extended}),\\
    & B'_\textrm{max}>2 \Rightarrow E<-2(1-\lambda)-\mu \cup 2(1-\lambda)-\mu<E\\
    & (\textrm{localized}).
\end{split}
\end{equation}

However, the result given by Eq.~(\ref{tb91}) is imperfect. In fact, it is only applicable for $0<\mu<2\lambda$ by the numerical verification, while for $2\lambda<\mu<2$ the upper mobility edge $2(1-\lambda)-\mu$ remains unchanged and the lower mobility edge becomes $-2(1+\lambda)+\mu$. Thus the  varying trend of the mobility edge becomes nonsmooth at the specific point $\mu=2\lambda$. We note that the similar singular effect has been
predicted~\cite{21pi} in the 3D speckle disorder model due to peculiarities of the potential correlation function. However, Ref.~\cite{22de} points out that this singularity is smoothed out by carrying the exact numerical calculation. For this model the nonsmooth change of the mobility edge is novel and definitive.

We have not found an exact analysis for these mobility edges, here we provide a naive picture to explain this phenomenon. We can make the proper deduction that when the strength of the on-site potential $\mu_{i}$ becomes very large, all wave functions of the system should be localized and the mobility edges must vanish due to the on-site potential being incommensurate. When the system reaches the all-wave-function-localized transition point, these two mobility edges must intersect at one point. However, the results given by Eq.~(\ref{tb91}) never intersect. Therefore, with the amplitude increments of $\mu_{i}$, the slope of the lower mobility edge must change sign, i.e., $-2(1-\lambda)-\mu\Rightarrow-2(1-\lambda)+\mu$, if the upper mobility edge remains unchanged. However, this change is discontinuous, and the continuous varying condition of the lower mobility edge requires $-2(1-\lambda)+\mu\Rightarrow-2(1+\lambda)+\mu$. Thus we can get $\mu=2\lambda$ from $-2(1-\lambda)-\mu =-2(1+\lambda)+\mu$, and $\mu=2$ from $2(1-\lambda)-\mu =-2(1+\lambda)+\mu$, and two mobility edges intersect at $\mu=2$, as shown in Fig.~\ref{0012}. In addition, for the existence of the mobility edges there is also an implicit condition such that $2\lambda<2$, i.e., $\lambda<1$. We summarize the expressions of mobility edges as follow:
\begin{equation}
    E_{\pm c2}=
    \left\{
                 \begin{array}{lr}
                 \pm 2(1-\lambda)-\mu,(0\leq\mu\leq2\lambda) &  \\
                 2(1-\lambda)-\mu, -2(1+\lambda)+\mu,(2\lambda<\mu<2) .
                 \end{array}
    \right.
\end{equation}

Thus, the model defined by the Hamiltonian~(\ref{tb1}) with
$\beta=(\sqrt{5}-1)/2$ has extended states at the band center ($E_{-c2}<E<E_{+c2}$) and localized states at the band edges ($E>E_{+c2}$, $E<E_{-c2}$), and the all-wave-function-localized transition point is located at $\mu=2$.

\section{Numerical verification}
\label{n3}
\begin{figure}
  \centering
  \includegraphics[width=0.5\textwidth]{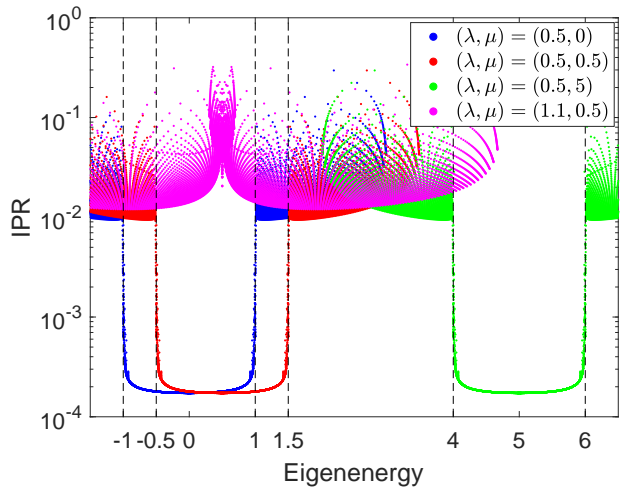}\\
  \caption{(Color online) The distribution of IPR as a function of eigenenergy for various
	$(\lambda,\mu)$ when $\beta=0$. ``Black dotted lines" correspond to two turning points of IPR located at the mobility edges $E_{\pm c1}=\pm2(1-\lambda)+\mu$. When $(\lambda,\mu)= (0.5,0), (0.5,0.5)$ and $(0.5,5)$, $E_{+c1}=1,1.5,6$ and $E_{-c1}=-1,-0.5,4$ are located at the spectrum respectively, while when $(\lambda,\mu)= (1.1,0.5)$, there are no mobility edges and all wave-functions are localized due to $\lambda=1.1>1$. }
  \label{0021}
\end{figure}
\begin{figure}
  \centering
  \includegraphics[width=0.5\textwidth]{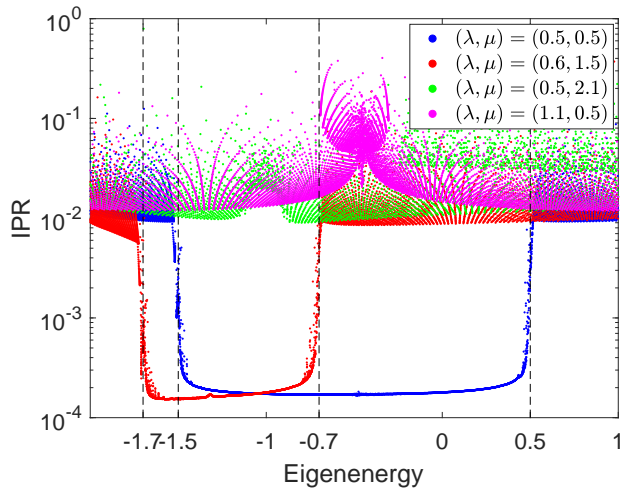}\\
  \caption{(Color online) The distribution of IPR as a function of eigenenergy for various
	$(\lambda,\mu)$ when $\beta=(\sqrt{5}-1)/2$. ``Black dotted lines" correspond to two turning points of IPR located at the mobility edges denoted by $E_{\pm c2}=\pm2(1-\lambda)-\mu$ if $0\leq\mu\leq2\lambda$ or by $E_{+c2}=2(1-\lambda)-\mu$ and $E_{-c2}=-2(1+\lambda)+\mu$ if $2\lambda<\mu<2$. When $(\lambda,\mu)= (0.5,0.5)$ and $(0.6,1.5)$, $E_{+c2}=0.5,-0.7$ and $E_{-c2}=-1.5,-1.7$ are located at the spectrum respectively, while when $(\lambda,\mu)= (0.5,2.1)$ and $(1.1,0.5)$, there are no mobility edges and all wave-functions are localized due to $\mu=2.1>2$ and $\lambda=1.1>1$ respectively.}
  \label{0022}
\end{figure}

 To support the semi-analytical results given in the previous section, we now present detailed numerical calculation. We diagonalize the model Hamiltonian (\ref{tb1}) directly to get the eigenenergies $E$ and the associated wave-functions $\psi$. Then the typical physical quantities used in the disordered system, such as the inverse participation ratio, the density of states and the Lyapunov exponent, can be obtained
 to distinguish the localized and extended states in the spectrum.

First we calculate the inverse participation ratio (IPR). The IPR of a normalized wave function $\psi$ is defined as~\cite{IPR1,IPR2,IPR3},
\begin{equation}
\text{IPR}_n =\sum_{j=1}^{L} \left|\psi^n_{j}\right|^{4},
\end{equation}
where $L$ denotes the total number of sites and $n$ is the index of energy level.
It is well known that the IPR of the extended state scales like $L^{-1}$, which approaches $0$ in the thermodynamic limit, but finite for a localized state.

Figure~\ref{0021} and Figure~\ref{0022} plot the IPR of the corresponding wave functions as a function of eigenenergy for various
$(\lambda,\mu)$ when $\beta=0$  and $(\sqrt{5}-1)/2$ respectively. We find that as the eigenenergy varies,
the IPR changes dramatically from the order of magnitude $10^{-2}$ (a typical value for the localized states) to $10^{-4}$ (a typical value for the extended states) or inversely at certain energies.
This jumping phenomenon of IPR indicates that there exist mobility edges in the energy spectrum. We implement calculations for various $(\lambda,\mu)$ and find that these mobility edges are exactly located at $E_{\pm c1}$ and $E_{\pm c2}$ respectively as analyzed in Sec.~\ref{n2}.

While if $\lambda=1.1$ the IPR of all wave-functions in both cases are of the order of magnitude of $10^{-2}$ and localized, and none of them appears around $10^{-4}$ and extended, which verifies the implicit condition $\lambda<1$ for the existence of mobility edges.
Remarkably, in Fig.~\ref{0021}, even if the strength of the potential ($\mu = 5$ in this situation) becomes very large, mobility edges $E_{\pm c1}$ still exist. Whereas in Fig.~\ref{0022}, when the strength of the slowly varying incommensurate potential is larger than the threshold ($\mu = 2$), all wave functions are localized and there are no mobility edges in the spectrum. As a result, a metal-insulator transition appears at $\mu = 2$ if $\beta=(\sqrt{5}-1)/2$, which is different from the $\beta=0$ case. We also choose different sets of parameters to ensure that mobility edges in the spectrum are indeed located at $E_{\pm c1}$ and $E_{\pm c2}$.

To strengthen the validity of the results of IPR, we also calculate the density of states $D(E)$ and the Lyapunov exponent $\gamma(E)$ of this quaisiperiodic system. Here $D(E)$ is defined as,
\begin{equation}
D(E) =\sum_{n=1}^{L} \delta(E-E_n),
\end{equation}
and $\gamma(E)$~\cite{14sarma} is,
\begin{equation}
\gamma(E_n) =\frac{1}{L-1}\sum_{n\neq m}^{L}\ln| E_n-E_m|,
\end{equation}
where $E_n$ is the $n$-th eigenenergy and $L$ is the total number of sites. The Lyapunov exponent is defined as the inverse localization length, hence $\gamma=0$ for an extended state whereas $\gamma \neq0$ for a localized state. The density of states and the Lyapunov exponent are connected by the relation
\begin{equation}
\gamma(E) =\int dE'D(E') \ln| E- E'|.
\label{tb13}
\end{equation}

\begin{figure}
  \centering
  \includegraphics[width=0.5\textwidth]{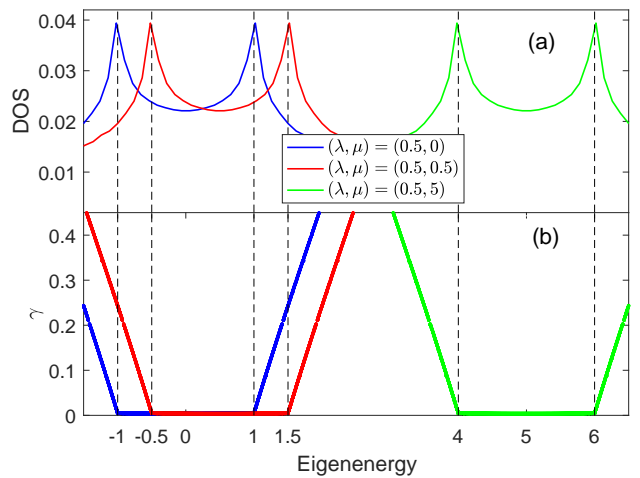}\\
  \caption{(Color online) DOS and the Lyapunov exponent $\gamma$ as a function of eigenenergy for various
	$(\lambda,\mu)$ when $\beta=0$. (a) Obviously a dramatic change of DOS occurs when the eigenenergy passes through the mobility edges $E_{\pm c1}$, which are in accordance with the IPR predictions. (b) When the eigenenergy is located in the interval $[E_{- c1}, E_{+ c1}]$, $\gamma \rightarrow 0$, indicating that the corresponding state is extended. Otherwise $\gamma$ is finite, indicating that the corresponding state is localized.}
  \label{0031}
\end{figure}

In Fig.~\ref{0031} we plot the density of states (DOS) and the Lyapunov exponent $\gamma$ as a function of the eigenenergy when $\beta=0$. To avoid losing generality, we choose three different sets of parameters with $(\lambda,\mu)=(0.5,0)$, $(0.5,0.5)$ and $(0.5,5)$. In Fig.~\ref{0031}(a) it is obviously that the DOS in our model displays sharp peaks at certain energies, which is different from 3D Anderson random disorder~\cite{gri,thou1}. The reason is due to the nature of the slowly varying incommensurate modulation~\cite{14sarma}, and this peak-like singularity of the DOS reflects
the change of the nature of the eigenstates. Therefore the extended-localized transitions corresponding to two mobility edges $E_{\pm c1}$ are indicated by two sharp peaks of the DOS. In Fig.~\ref{0031}(b) $\gamma$ is plotted for the same sets of parameters, it also exhibits singular behaviors at two mobility edges, just as those for the DOS.

\begin{figure}
  \centering
  \includegraphics[width=0.5\textwidth]{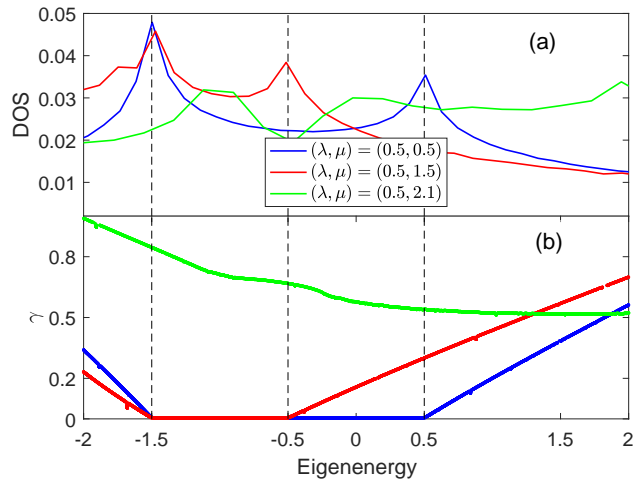}\\
  \caption{(Color online) DOS and the Lyapunov exponent $\gamma$ as a function of eigenenergy for various
	$(\lambda,\mu)$ when $\beta=(\sqrt{5}-1)/2$. (a) Obviously a dramatic change of DOS occurs when the eigenenergy passes through the mobility edges $E_{\pm c2}$ when $\mu<2$, while when $\mu=2.1>2$, there is no obvious change. (b) When $\mu<2$ $\gamma \rightarrow 0$ for the the eigenenergy located in the interval $[E_{- c2}, E_{+ c2}]$, indicating that the corresponding state is extended, otherwise $\gamma$ is finite, indicating that the corresponding state is localized. Whereas when $\mu=2.1>2$, the $\gamma$ is finite with eigenenergy varying, indicating that all wave functions are localized.}
  \label{0032}
\end{figure}

In Fig.~\ref{0032} we show the numerical results for the DOS (Fig.~\ref{0032}(a)) and $\gamma$ (Fig.~\ref{0032}(b)) when $\beta=(\sqrt{5}-1)/2$. They also exhibit the similar singularity as the eigenenergy passes through two mobility edges $E_{\pm c2}$ for $\mu<2$.
When $(\lambda,\mu)=(0.5,2.1)$, the DOS and $\gamma$ vary smoothly and do not exhibit any singular behavior, so there exist no mobility edges and all wave functions are localized compared with the $\mu<2$ case, this agrees with the semi-analytical prediction given in Sec.~\ref{n2}.
We have also checked other combinations of parameters and get the same results as expected.

Another interesting subject is the critical exponents of the Lyapunov exponent at the mobility edge defined by
\begin{equation}
\gamma(E) \sim | E- E_{\pm c i}|^\theta,\, i=1,2,\,
\end{equation}
in the localized regions of energy spectrum. Similarly, the behavior of the density of states at the mobility edge can be written as
\begin{equation}
D(E) \sim | E- E_{\pm c i}|^{-\delta},\, i=1,2.
\end{equation}
Therefore, according to Eq.~(\ref{tb13}) the critical exponents $\theta$ and $\delta$ are clearly related by the equation
\begin{equation}
\theta + \delta =1.
\end{equation}

In Fig.~\ref{0031}(b) and Fig.~\ref{0032}(b), we can identify that
 $\gamma$ behaves to be linear with eigenenergy varying in the localized region, i.e., the band edges ($E>E_{+c1,2}$, $E<E_{-c1,2}$), which leads to the fact that $\theta=1$ and $\delta=0$. These results agree with those of the known model~\cite{14sarma}, and the parameters $\lambda$, $\mu$ and $v$ are all found to be irrelevant to the critical exponents $\theta$ and $\delta$. In addition, we also find that
these mobility edges depend on $\lambda$ and $\mu$ but are irrelevant to $v$ by varying the parameters.

\section{Conclusions}
\label{n4}
In this work we have studied the localization properties of a class of the off-diagonal disordered tight-binding models with a constant on-site potential ($\beta=0$) and a slowly varying incommensurate on-site potential ($\beta=(\sqrt{5}-1)/2$), we find following interesting features of two models.

(1) When $\beta=0$, we reveal that there exist two mobility edges $E_{\pm c1}=\pm2(1-\lambda)+\mu$ separating localized and extended states
states in the spectrum. More interestingly, this phenomenon completely results from the off-diagonal disorder term, and even if the strength of the on-site potential $\mu$ becomes very large, mobility edges $E_{\pm c1}$ still exist.

(2) When $\beta=(\sqrt{5}-1)/2$, there also exist two mobility edges $E_{\pm c2}$ in the spectrum. However, when the incommensurate potential strength satisfies $0\leq\mu\leq2\lambda$ we have $E_{\pm c2}=\pm 2(1-\lambda)-\mu$, whereas when $2\lambda<\mu<2$ we have $E_{+ c2}= 2(1-\lambda)-\mu$ and $ E_{- c2}= -2(1+\lambda)+\mu$. When $\mu>2$ all wave-functions of the model are localized.

In summary, we show that the semi-analytical critical conditions of mobility edges are in excellent agreement with the localization properties obtained from the numerical calculation, we believe that the interesting features of these models will bring new perspectives to a wide range of correlated disordered systems.

\begin{acknowledgments}
G. H. thanks the support from the NSF of China (Grant No.
11674051).

\end{acknowledgments}

\end{document}